\documentclass[11pt]{article}
\usepackage{amssymb}
\usepackage{amstext}
\usepackage{amsmath}
\usepackage{latexsym}
\usepackage{color}
\usepackage{natbib}
\usepackage{indentfirst}

\usepackage{verbatim}

\newcommand{\RR}{\mathbb R}

 \newcommand{\bea}{\begin{eqnarray}}
 \newcommand{\eea}{\end{eqnarray}}
 \newcommand{\bmat}{\begin{pmatrix}}
 \newcommand{\emat}{\end{pmatrix}}
 \newcommand{\x}{{\bf x}}
 \newcommand{\y}{{\bf y}}
\newcommand{\p}{{\bf p}}
\newcommand{\q}{{\bf q}}
\newcommand{\kv}{{\bf k}}

\newtheorem{thm}{Theorem}

\newtheorem{defn}[thm]{Definition}

\newtheorem{rem}[thm]{Remark}

\numberwithin{equation}{section}

\title{Coadjoint Orbits of the Poincar\'e Group in $2+1$ Dimensions and their Coherent States}
\author{V. Hudon \thanks{Department of Mathematics, University of Tennessee, Knoxville, Tennessee, USA 37996} \and S.Twareque Ali  \thanks{Department of Mathematics and Statistics, Concordia University, Montr\'eal, Qu\'ebec, CANADA H3G 1M8}}
\date{\today}

\begin{document}
\bibliographystyle{plainnat}

\maketitle

\begin{abstract}
\noindent
We study the structure of the coadjoint orbits of the $2+1$ Poincar\'e group, using a matricial representation of the group. We also obtain the orbits connected to irreducible representations of the group. Finally we obtain coherent states for the hyperboloidal and conical orbits. 
\end{abstract}

\section*{Introduction}

Three-dimensional space-times have recently been studied in connection with general relativity, black holes and gravitation (see, for example, \citet{bateza,giabku,wit}). For such studies the Poincar\'e group in 2-space and 1-time dimensions is the symmetry group of the underlying geometry. In the area of signal processing, the analysis of two-dimensional signals, varying with time, is a major preoccupation in current day research. Again, a group such as the Poincar\'e group in $2+1$ dimensions is a representative candidate for the construction of relevant signal transforms. For all these reasons, an analysis of this group and a construction of its coherent states is an important mathematical problem.

In this paper, we study the Poincar\'e group in $2+1$ dimensions by analyzing its coadjoint orbits structure.  In the first part, we explicitly compute the coadjoint action, using a matricial representation of the group and we then classify the coadjoint orbits.  In the second part, we define coordinates and invariant measures on the upper sheet of the two-sheeted hyperboloid and the upper cone in order to, first, induce a unitary irreducible representation and, second, build coherent states on them.

The Poincar\'e group is the symmetry group of $(2+1)$-dimensional relativity.  It is written as a semidirect product ${\mathbb R}^{n,1} \rtimes SO(n,1)$.  Its well-known $(3+1)$-dimensional version has been studied in great detail in the literature (see, for example, \citet{KimNoz}).  The $(2+1)$-dimensional version has recently been studied in \citet{Gitman1997_poincare21dim}, where a unitary irreducible representation has been constructed and the orbits of the action of $SO(2,1)$ on ${\mathbb R}^{2,1}$ have been obtained, using wave equations.

The orbit structure of the full Poincar\'e group, related its irreducible representations, have been extensively studied in the literature, the earliest being due to \citet{wig}, reproduced also in \citet{BarutR,KimNoz}. The method is  based on fixing a  momentum vector in $3+1$ dimensions, looking at its stability subgroup and its orbit under the Lorentz group. One obtains a one-sheeted hyperboloid when the rest mass $m^2 < 0$, a cone for $m^2 = 0$ (upper and lower for $p_0 > 0$, $p_0 < 0$), the origin (degenerate orbit) when $m^2 =0$ and $p_0 = 0$,  and a two-sheeted hyperboloid for $m^2 > 0$ (upper and lower for $p_0 > 0$, $p_0 < 0$). These lattter are three-dimensional orbits.

In the $(2+1)$-dimensional case, in \citet{AlmoroxPrieto2001_LHhyp} the authors  study the action of $SO(2,1)$ on ${\mathbb R}^{2,1}$ and by an analysis using the moment map, they  obtain similar two-dimensional orbits.
In this paper, we work with a matricial representation of the group and obtain directly the orbits by the action of the $SO(2,1)$ group matrices on some chosen vectors.  Moreover, we compute and classify
 the coadjoint orbits of the group which give its unitary irreducible representations.  The relation between the orbits and the representations is  outlined in \citet{Kirillov,Kirillov2,Kirillov1999_orbmeth_merits}, where it is also shown  that the coadjoint orbits are symplectic manifolds.
In this paper, we also link some of the coadjoint orbits to the orbits in $\RR ^{2,1}$ under the action of $SO(2,1)$.

Next, following \citet{Mackey}, we compute the induced representations related to two of the coadjoint orbits (the hyperboloid and the cone).  (The inducing technique allows one to obtain the representation of a group from a known representation of one of its subgroups.)  In this paper, we are using a representation of the stabilizer as a starting point in the inducing technique.  This will give us a convenient form of the representation for building coherent states.

  Coherent states were first developed in the context of quantum optics.  The idea has seen explosive development in many areas of mathematics and physics.
In the group context, coherent states are generally obtained by the action of a unitary representation on a specific set of vectors.  The technique is discussed in detail, in for example, \citet{AliAG,Antoineetal}, where applications to signal analysis are also discussed.  A more recent and useful example of this technique is given in \citet{AntoineMahara_1999_CSaffGalgr}, where the authors compute the orbits and coherent states of the affine Galilei group.

The literature on the coherent states for the Poincar\'e group in $1+1$ and $3+1$ dimensions (see \citet{AliAG} and references therein) is extensive.  As already stated, here we focus on the $(2+1)$-dimensional case which is less covered in the literature.
In \citet{Gitman1997_poincare21dim} the authors obtain  coherent states for $SU(1,1)$ (which is isomorphic to $SO(2,1)$) in the framework of harmonic analysis.
In \citet{deBievre1989_CSsympl}  coherent states of semidirect product groups through orbits are obtained using the moment map. In \citet{Bohnke1991_coneCS}  tight frames are constructed using coherent states on the forward light cone, using the $(n+1)$-dimensional Poincar\'e group with  dilation. Our focus here is on the specific group mentioned and  to perform a complete and  full computation in an explicit way and to obtain the coherent states on the hyperboloid and the cone following the technique described in \citet{AliAG}.

To summarize, we first obtain a formula for the coadjoint action of the Poincar\'e group $G = \mathbb{R}^{2,1} \rtimes SO(2,1)$ in a concrete matrix representation.  It is given in \eqref{coad_act} and uses a particular representation of the algebra.
We then compute explicitly the orbits of $SO(2,1)$ on $\RR ^{2,1}$ and the coadjoint orbits.
We obtain a degenerate orbit, the upper and lower sheets of a two-sheeted hyperboloid, the upper and lower cone and the one-sheeted hyperboloid.  The hyperboloids and the cones appear as our  first orbits and also as two-dimensional coadjoint orbits.  They also appear together with their cotangent spaces as four-dimensional coadjoint orbits.
We define coordinates on the orbits and compute invariant measures on them in order to obtain representations based on the hyperboloid and the cone.  Next, we compute the coherent states on both of them.
For the hyperboloid, we obtain coherent states using the so-called principal section.
For the cone, we obtain a family of coherent states for a generalized principal section.

The rest of this paper is divided as follows.  In Section \ref{sect_groupaction}, we first describe the group, especially the representation of the algebra we will be using.  We also compute the expression for the coadjoint action and give a useful bijection.
In Section \ref{sect_orbits}, we compute explicitly all types of orbits for different initial vectors and study the link among the different orbits obtained.
Then, in Section \ref{meth_comput}, we give some definitions and present the technique used to obtain the coherent states.  This technique is then applied to the hyperboloid in Section \ref{cs_hyp} and the cone in Section \ref{cs_cone}.

\section{Description of the group and actions}\label{sect_groupaction}

We present here the definitions needed to compute the orbits.  We first give a description of the group and a particular basis for the algebra.  After that we compute the adjoint and coadjoint actions in this basis.  Finally, we give the definitions of the representation generating orbit and of a certain bijection.

\subsection{Matricial representation}

We are working in a space-time of $2+1$ dimensions.
The Poincar\'e group is written as a semidirect product group:
$G = {\mathbb R}^{2,1} \rtimes SO(2,1)$.
An element is written:
\bea \label{group_el}  G \ni g = \bmat \Lambda & v \\ 0 & 1 \emat, \eea
where $\Lambda \in SO(2,1)$ and $v \in {\mathbb R}^{2,1}$.
$\Lambda \eta \Lambda ^{\dagger} = \eta$, $\eta$
being the metric: ${\rm diag}(+1, -1, -1)$.
The six algebra generators of the Poincar\'e group are given by:
\bea \label{alggen}
J_0 = \bmat 0 & 0 & 0 \\ 0 & 0 & -1 \\ 0 & 1 & 0 \emat ,
J_1 = \bmat 0 & 0 & 1 \\ 0 & 0 & 0 \\ 1 & 0 & 0 \emat ,
J_2 = \bmat 0 & 1 & 0 \\ 1 & 0 & 0 \\ 0 & 0 & 0 \emat , \nonumber \\
P_0 = \bmat 1 \\ 0 \\ 0 \emat,
P_1 = \bmat 0 \\ 1 \\ 0 \emat,
P_2 = \bmat 0 \\ 0 \\ 1 \emat.
\eea
We also define $J_+ = J_0 + J_1$ and $J_- = J_0 - J_1$.  They exponentiate to translations (see \eqref{lambda_translation}).

Here are the one-parameter subgroups of $SO(2,1)$:
{\footnotesize \bea \label{group_basis}
\Lambda _{J_0} = \bmat 1 & 0 & 0 \\ 0 & \cos \alpha & -\sin \alpha \\ 0 & \sin \alpha & \cos \alpha \emat,
\Lambda _{J_1} = \bmat \cosh \beta & 0 & \sinh \beta \\ 0 & 1 & 0 \\ \sinh \beta & 0 & \cosh \beta \emat,
\Lambda _{J_2} = \bmat \cosh \gamma & \sinh \gamma & 0 \\ \sinh \gamma & \cosh \gamma & 0 \\ 0 & 0 & 1 \emat,
\eea }
\begin{gather} \label{lambda_translation}
\Lambda _{J_+} = \bmat 1 + \frac{u^2}{2} & \frac{u^2}{2} &  u \\  -\frac{u^2}{2} & 1- \frac{u^2}{2} &  -u \\  u & u & 1  \emat, \ \
\Lambda _{J_-} = \bmat  1+\frac{v^2}{2}  & \frac{-v^2}{2}   & -v   \\  \frac{v^2}{2}  &  1-\frac{v^2}{2}  &  -v  \\  -v  & v   &  1    \emat.
\end{gather}
The $\Lambda _{J_{\pm}}$ matrices will be used for obtaining the conical orbit in Section \ref{sect_cone_orbit}.

The Iwasawa decomposition of the $SO(2,1)$ group will also be useful in the conical case.
We write an element of $SO(2,1)$ as the product of three elements, so $g=kan$.  Where we have that $k$ is a rotation which corresponds to $\Lambda _{J_0}$, $a$ is a boost (or dilation) which corresponds to $\Lambda _{J_2}$ and $n$ is a translation which corresponds to $\Lambda _{J_{\pm}}$ depending on the case.
Instead of being $J_0$, $J_1$ and $J_2$, the algebra generators are now taken to be $J_0$, $J_2$ and $J_0 \pm J_1$.

\subsection{$(\alpha, \beta)$ basis for the algebra}

Any element of the Lie algebra can be written as a linear combination:
\bea \label{mat_alg_ab} X = \bmat \alpha ^t \cdot J  & \beta \\ 0 & 0 \emat, \eea
where $\alpha$ and $\beta$ are three-column vectors, $J$ is the vector $(J_0,J_1,J_2)^t$ and the product $\cdot$ is
simply the linear combination:
$ \alpha ^t \cdot J = \alpha_0 J_0 +\alpha_1 J_1 + \alpha_2 J_2$.
To compute the action, it will be easier to work in the six parameters space of $\alpha$ and $\beta$.  We thus rewrite the element $X$ as a column vector $X = \bmat \alpha \\ \beta \emat$.
Elements in the dual algebra are then written as row vectors:
\bea
X = \bmat \alpha \\ \beta \emat \to X^* = \bmat \alpha ^* & \beta ^* \emat,
\eea
where $\alpha ^*$ and $\beta ^*$ are themselves three-dimensional row vectors.
The dual pairing in this notation is simply the scalar product, we choose not to use the metric in this setting.

\subsection{Adjoint and coadjoint action}

The adjoint action is the action of a group $G$ on its own algebra $\mathfrak g$.
For a matrix group, it is defined by:
\bea \label{def_adjac}
Ad (g) X  =  g X g^{-1} ,
\eea
where $g \in G$ and $X \in \mathfrak g$.
The coadjoint action is the action of the group on the dual of its algebra.
Generally, the coadjoint action (denoted $Ad^\#$) is defined as in:
\bea \label{def_coadact}
<Ad^\# (g) X_1^* , X_2> = <X_1^* , Ad(g^{-1})X_2>.
\eea

\subsubsection{Adjoint action for the Poincar\'e group}

Using the definition \eqref{def_adjac} and the algebra element in the $(\alpha, \beta)$ basis \eqref{mat_alg_ab},
the adjoint action of the group $G$ on its algebra is given by:
\bea
Ad (g) X & = & g X g^{-1} \nonumber \\
 & = & \bmat  \Lambda \alpha ^t \cdot J \Lambda ^{-1} &  -\Lambda \alpha \cdot J \Lambda ^{-1}v + \Lambda \beta
\\ 0 & 0 \emat  \nonumber \\
 & \equiv & \bmat  \alpha '^t \cdot J & \beta ' \\ 0 & 0 \emat.
\eea
We compute the part $\Lambda \alpha ^t \cdot J \Lambda ^{-1}$
for a generic element of the group
$\Lambda = \Lambda _{J_0} \Lambda _{J_1} \Lambda _{J_2}$
in order to extract the linear combination of $J$'s.  After a few manipulations, we get:
\bea \label{rewrite_adj1}
\Lambda \alpha ^t \cdot J \Lambda ^{-1} = (m (\Lambda ^{-1})^t m \alpha)^t \cdot J ,
\eea
where the matrix $m$ is:
\bea  m =  \bmat 1 & 0 & 0 \\ 0 & 1 & 0 \\ 0 & 0 & -1 \emat. \eea
%
%
%
For convenience, we set the following notation: $m (\Lambda ^{-1})^t m \equiv \Hat{\Lambda}^{-1}$ and, then, $\Hat{\Lambda} = m \Lambda ^t m$.
This is an inner automorphism of the group.
%

We can also work out the following by direct computations:
\bea \label{rewrite_adj2}
-\Lambda \alpha \cdot J \Lambda ^{-1}v = -(J\cdot v) \Hat{\Lambda}^{-1} \alpha ,
\eea
where $J\cdot v$ is the matrix $(J_0 v, J_1 v , J_2 v)$, recalling that $v$ is a three-column vector.
We write it down explicitly here for later use:
\bea \label{Jv}
J\cdot v = \bmat  0 & v_2 & v_1 \\ -v_2  & 0 & v_0 \\ v_1 & v_0 & 0  \emat.
\eea

We have thus obtained the transformation of the parameters $\alpha$ and $\beta$:
\bea
\alpha ' \cdot J &=& \Lambda \alpha \cdot J \Lambda ^{-1} \nonumber \\
& =& \Hat{\Lambda} ^{-1} \alpha \cdot J \nonumber \\
\beta ' &=& -\Lambda \alpha \cdot J \Lambda ^{-1}v + \Lambda \beta \nonumber \\
& =& \Lambda \beta - (J\cdot v) \Hat{\Lambda} ^{-1} \alpha . \nonumber
\eea
We can rewrite the transformation as a $6 \times 6$ matrix $M(g)$:
\bea \label{matM}
\bmat \alpha ' \\ \beta ' \emat = M(g) \bmat \alpha \\ \beta \emat \equiv
\bmat \Hat{\Lambda} ^{-1}  &  0  \\  - (J\cdot v) \Hat{\Lambda} ^{-1}  &  \Lambda   \emat \bmat \alpha \\ \beta \emat .
\eea
The adjoint action of $g = (\Lambda, v)$ on $X = (\alpha, \beta)^t$ is then written in a matricial form:
\bea Ad (g) X = M(g) X. \nonumber \eea

\subsubsection{Coadjoint action for the Poincar\'e group}

We now define and compute the coadjoint action in the six parameters space of $\alpha ^*$ and $\beta ^*$.
In this notation, the equation \eqref{def_coadact} reads:
\bea
<X_1^* , Ad(g^{-1})X_2> = X_1^* M(g^{-1})X_2 = <Ad^\# (g) X_1^* , X_2>,
\eea
where $M(g)$ is defined in (\ref{matM}).
The coadjoint action is then:
\bea
Ad^\# (g) X^* = \bmat \alpha ^* & \beta ^* \emat M (g^{-1}).
\eea
The matrix $M(g^{-1})$ is easily obtained from $M(g)$ using the inverse of a group element $g^{-1}= (\Lambda ^{-1}, -\Lambda ^{-1} v)$:
\bea
M (g^{-1}) = \bmat \Hat{\Lambda}  &  0  \\   (J\cdot (\Lambda ^{-1} v)) \Hat{\Lambda}  &  \Lambda ^{-1} \emat .
\eea
We also compute the inverse of the matrix $M(g)$:
\bea
M (g)^{-1} = \bmat \Hat{\Lambda}  &  0  \\  \Lambda ^{-1}(J \cdot v)   &  \Lambda ^{-1} \emat .
\eea
By direct computation for the different one-parameter subgroups, we get that
\bea \label{rewrite_coad} (J\cdot (\Lambda ^{-1} v)) \Hat{\Lambda} = \Lambda^{-1}(J \cdot v).\eea
  Then, as one would expect, $M (g^{-1})$ and $M (g)^{-1}$ are the same.  We will use $M (g)^{-1}$ to define the coadjoint action.
This gives:
\begin{equation}\label{coad_act}
   Ad^\# (g) X^* = \bmat \alpha ^* \Hat{\Lambda} + \beta ^* \Lambda ^{-1}(J\cdot v)  &  \beta ^* \Lambda ^{-1} \emat ,
   \end{equation}
where $\Lambda$ and $v$ are fixed by the choice of $g$, an element of the group.  The choice of $X^*$ varies for the different cases.

\subsection{Representation generating orbits}

For a semidirect product group $G= V \rtimes S$, the orbits of the action of $S$ on $V^*$ will be called {\it representation generating orbits} since those orbits are used to obtain representations in the induced representation method.

To get the action of the $SO(2,1)$ part of the Poincar\'e group on the dual of ${\mathbb R}^{2,1}$,
we just multiply the row-vector $X^* = \bmat \gamma _0 & \gamma _1 & \gamma _2  \emat $ by the subgroup matrices given in \eqref{group_basis}.

\subsection{Bijections}

There is a bijection between the orbit of a point $x$ and the
quotient of the group $G$ by the stabilizer of this point: $\mathcal{O} _x \simeq G / H_x$.
This bijection is used to obtain our coadjoint orbits.

In the case of a semidirect product group, there is also an isomorphism relating some
coadjoint orbits and the orbits of the action of $S$ on $V^*$ which is given in \citet{AliAG}.  It is based on the following facts and conventions:
\begin{itemize}
\item the group is $G = V \rtimes S$, $V$ a vector space, $S \subset GL(V)$;
\item $H_0$ denotes the stabilizer of $(0,k_0) \in \mathfrak{g}^*$ under the coadjoint action,
 $k_0 \in V^*$;

\item $\mathcal{O} _{(0,k_0)}$ is the orbit of $(0,k_0)\in \mathfrak{g}^*$ under the coadjoint action of $G$;
\item $T^*\mathcal{O}^*$ is the cotangent bundle of the orbit of $k_0$ in $V^*$ under $S$.
\end{itemize}
The equation (10.49) in \citet{AliAG} gives the following sequence of isomorphisms:
\bea \label{isom}
\Gamma = G/H_0 \simeq
\mathcal{O} _{(0,k_0)}
\simeq T^*\mathcal{O}^*.
\eea
We will explicitly check this isomorphism for some particular vectors $k_0$ later.
We will see that $\mathcal{O} _{(0,k_0)}$ are the four-dimensional coadjoint orbits.

\section{Orbits}\label{sect_orbits}

We now present the orbits obtained from different initial vectors.  Those initial vectors have been chosen to cover different cases: a purely time-like vector ($k_0 = (\pm m,0,0)$), a purely space-like vector ($k_0 = (0,m,0)$) and a mixed time-space-like vector ($k_0 = (\pm 1,1,0)$).  We also cover the trivial case $k_0 = (0,0,0)$.

For each vector, we present the representation generating orbit and the coadjoint orbit obtained using \eqref{coad_act} with both $X^* = (0,k_0)$ and $X^* = (k_0,0)$.

\subsection{Degenerate orbit}

The first case is simple.  We start with the initial vector $k_0 = (0, 0, 0)$.  Since the vectors $(0, k_0)$ and $(k_0, 0)$ are the same, we only have one coadjoint orbit.

\paragraph{Representation generating orbit}

When we multiply $X^*=k_0= (0, 0, 0)$ by the subgroup matrices, we simply obtain the origin, that is a degenerate orbit.

\paragraph{Coadjoint orbit}

Using \eqref{coad_act} with the null vector, we again get only the point at the origin.
The orbit we obtain is then again degenerate.
In this case, the stabilizer is the whole group $G$.

\subsection{Two-sheeted hyperboloidal orbit}\label{sect_2sheethyp_orbit}

We move to a more interesting case, we study the orbits emerging from the initial vector $k_0 = (\pm m, 0, 0)$, where $m>0$ can be seen as the mass.

\paragraph{Representation generating orbit}

We multiply the row-vector $k_0$ by the subgroup matrices given in \eqref{group_basis}.
The two boosts are acting giving two hyperbolas.  This generates the two-sheeted hyperboloid.

For $k_0 = (m, 0, 0)$, we get the upper sheet of the two-sheeted hyperboloid with its vertex at $q_0=m$.
For $k_0 = (-m, 0, 0)$, we get the lower sheet of the two-sheeted hyperboloid with its vertex at $q_0=-m$.

\paragraph{Four-dimensional coadjoint orbit}

We use \eqref{coad_act} with the vector $(\alpha ^*, \beta ^*) = (0, k_0) = (0, 0, 0, \pm m, 0, 0)$.
The stabilizer $H_0$ consists of the rotation and the time translation.
The quotient of the group by the stabilizer leaves the two boosts
and the two space translations to generate the coadjoint orbit.
We thus have the upper or lower sheet of the two-sheeted hyperboloid (depending on the sign in $\pm m$) together with the spatial plane, that is a four-dimensional coadjoint orbit.

The equation of the hyperboloid is $q_0^2-q_1^2-q_2^2 = m^2$.  The vertices are at $\pm m$ and the hyperboloids all have the same cone as an asymptote.

\paragraph{Two-dimensional coadjoint orbit}

We now use \eqref{coad_act} with $(\alpha ^*, \beta ^*) = (k_0, 0) = (\pm m, 0, 0, 0, 0, 0)$.
The stabilizer is made up of the rotation and the three translations.
We are left only with the two boosts to generate the coadjoint orbit.
We thus get the upper and lower sheet of a two-sheeted hyperboloid with $m$ and $-m$ respectively.
This is different from the previous case since we are left with a two-dimensional structure instead of a four-dimensional one.

\subsection{Conical orbit}\label{sect_cone_orbit}

In this section, we study the orbits arising from the initial vector $k_0 = (\pm 1, 1, 0)$.
For this case, we need the Iwasawa decomposition and the translation matrix of $SO(2,1)$ given in \eqref{lambda_translation}.

\paragraph{Representation generating orbit}

Here are the results for $X^* =k_0 = \bmat \pm 1 & 1 & 0 \emat$:
\begin{subequations}
\bea
X^* \Lambda _{J_0} &=&
\bmat  \pm 1 &  \cos \alpha  & - \sin \alpha  \emat, \label{2a} \\
X^* \Lambda _{J_1} &=&
\bmat  \pm  \cosh \beta  & 1 & \pm \sinh \beta  \emat,  \label{2b} \\
X^* \Lambda _{J_2} &=&
\bmat  \pm \cosh \gamma +  \sinh \gamma & \pm \sinh \gamma + \cosh \gamma & 0 \emat . \label{2c}
\eea
\end{subequations}

From \eqref{2a}, we see that if we cut the orbit at $t=\pm 1$, we have a circle.
From \eqref{2b}, we see that if we cut the orbit at $x=1$, we have a hyperbola.
From \eqref{2c}, we see that if we cut the orbit at $y=0$, we have a straight line.
Moreover, in all cases, we have that $t^2 - x^2 - y^2=0$.

This thus gives the upper cone for the vector $k_0 = (1, 1, 0)$ and the lower cone for $k_0 = (-1, 1, 0)$.

\paragraph{Four-dimensional coadjoint orbit}

We use \eqref{coad_act} with the vector $(\alpha ^*, \beta ^*) = (0, k_0) = (0, 0, 0, \pm 1, 1, 0)$.
The stabilizer $H_0$ is made up of a translation $\Lambda _{J_{\pm}}$ (or $n$ in the Iwasawa decomposition) and the vector $x=\mp t$.
The orbit is then generated by the rotation and the boost ($k$ and $a$ in the Iwasawa decomposition) which gives the cone with the plane generated by the $y$-axis and the axis $x=\pm t$.  This is a four-dimensional orbit.

The cone equation is $q_0 ^2 - q_1^2 -q_2^2 =0$.
It is the upper cone for $q_0 > 0$ and the lower one for $q_0 < 0$.
This cone is actually the limiting case of the two-sheeted hyperboloid in the massless limit.

\paragraph{Two-dimensional coadjoint orbit}

We now use $(\alpha ^*, \beta ^*) = (k_0, 0) = (\pm 1, 1, 0, 0, 0, 0)$ as the initial vector in \eqref{coad_act}.
In this case, the vector is stabilized by the $n$ translation of $SO(2,1)$ and also by all of the $\RR ^{2,1}$ translations.
The quotient thus leaves the rotation and the boost to generate the orbit, a
two-dimensional cone.

\subsection{One-sheeted hyperboloidal orbit}

We now present the study of the orbits originating from the initial vector $k_0 = (0, m, 0)$.

\paragraph{Representation generating orbit}

We have the action of the rotation and one of the boosts.  This gives the one-sheeted hyperboloid.

The representation generating orbit is then the one-sheeted hyperboloid which cuts the spatial plane at the circle of radius $m$.

\paragraph{Four-dimensional coadjoint orbit}

Using the vector $(\alpha ^*, \beta ^*) = (0, k_0) = (0, 0, 0, 0, m, 0)$ in \eqref{coad_act}, we obtain that
the stabilizer is the boost in the $x$-direction ($\Lambda _{J_1}$) and the $x$-translation ($v_1$).
The orbit is then generated by the rotation and the $y$-boost to which we add the $t$- and $y$-translations.
Geometrically, this is the one-sheeted hyperboloid together with the $ty$-plane.
The equation of the one-sheeted hyperboloid is $q_0^2-q_1^2-q_2^2 = -m^2$.

\paragraph{Two-dimensional coadjoint orbit}

 On the other hand, if we use
 $X^* _0 = (\alpha ^*, \beta ^*) = (k_0, 0) = (0, m, 0, 0, 0, 0)$ in \eqref{coad_act},
the stabilizer is the boost in the $x$-direction together with the three translations.
We thus obtain the rotation and the $y$-boost as the orbit generators.  Geometrically, we can see it as a one-sheeted hyperboloid, hence a two-dimensional structure.

\subsection{Summary of the orbits and study of the isomorphisms}\label{sect_summary}

\begin{table}
\center{
\begin{tabular}{|c|c|c|c|}
\hline
Fixed vector &  Stabilizer  &  Orbit generators &  Geometry   \\ \hline  \hline
$(0, 0)$  & whole $G$  &  nothing  & origin (degenerate)  \\ \hline
$(k_0, 0)$  & $\Lambda _{J_0}$  &  $\Lambda _{J_1}$, $\Lambda _{J_2}$  & 2-sheet hyp.  \\
$k_0= (\pm m, 0, 0)$ & $v_0$, $v_1$, $v_2$  &   &  [upper (+) and lower (-)] \\  \hline
$(0, k_0)$  & $\Lambda _{J_0}$  &  $\Lambda _{J_1}$, $\Lambda _{J_2}$  & 2-sheet hyp. + $xy$-plane  \\
$ k_0=(\pm m, 0, 0)$ & $v_0$  & $v_1$, $v_2$  &  [upper (+) and lower (-)] \\  \hline
$(k_0, 0)$  & $\Lambda _{J_{\pm}}$  & $\Lambda _{J_0}$, $\Lambda _{J_2}$  & cone  \\
$k_0= (\pm 1, 1, 0)$ & $v_0$, $v_1$, $v_2$  &   &  [upper (+) and lower (-)] \\  \hline
$(0, k_0)$  & $\Lambda _{J_{\pm}}$  & $\Lambda _{J_0}$, $\Lambda _{J_2}$  & cone + plane  \\
$k_0=(\pm 1, 1, 0)$ & $x= \mp t$   &   $x= \pm t$, $v_2$  &  [upper (+) and lower (-)] \\  \hline
$(k_0, 0)$  & $\Lambda _{J_1}$  &  $\Lambda _{J_0}$, $\Lambda _{J_2}$  & 1-sheet hyp.  \\
$k_0=(0, m, 0)$  & $v_0$, $v_1$, $v_2$ &  &  \\ \hline
$(0, k_0)$  & $\Lambda _{J_1}$  &  $\Lambda _{J_0}$, $\Lambda _{J_2}$  & 1-sheet hyp. + $ty$-plane   \\
$k_0=(0, m, 0)$  & $v_1$  & $v_0$, $v_2$ & \\ \hline
\end{tabular}}
\caption{Coadjoint orbits of the group $G = {\mathbb R}^{2,1} \rtimes SO(2,1)$}
\label{sum_coad}
\end{table}

We present a recapitulation of the coadjoint orbits obtained in Table \ref{sum_coad}.
It is interesting to remark that both the one-sheeted and two-sheeted hyperboloids have the cone as an asymptote.
The initial vectors $k_0$ that we have used cover all the cases, that is a purely time, a purely space and a mixed time-space initial vector.
We retrieve the same representation generating orbits found in the literature.


It is possible to study the isomorphism \eqref{isom} for the cases presented above.
\bea
\Gamma = G/H_0 \underbrace{\simeq}_{1}  \mathcal{O} _{(0,k_0)}
\underbrace{\simeq}_{2} T^*\mathcal{O}^*,  \nonumber
\eea
The first isomorphism (1) has been used to compute the coadjoint orbit.  The other isomorphism (2) links the four-dimensional coadjoint orbit to the cotangent bundle of the representation generating orbit.
We easily can see that the four-dimensional orbits are the cotangent bundles of those orbits.
The isomorphism is thus verified.

\section{Coherent states: definitions and methods} \label{meth_comput}

In this section, we describe the induced representation method as it will be used in the following.
We also outline the method used to obtain the coherent states.

\subsection{Induced representations}\label{ind_rep_meth}

The method that we use to get representations for the computation of the coherent states in Sections \ref{cs_hyp} and \ref{cs_cone} follows \citet{AliAG}, \S 10.2.4.

Let $d\nu$ be the invariant measure on the four-dimensional coadjoint orbit $\Gamma$ and take
the corresponding  Hilbert space  $L^2(\Gamma, d\nu)$.

We first associate a unitary character $\chi$ to $V = \mathbb{R}^{2,1}$ in this way:
\bea
\chi (v) = \exp (-i<k_0;v>),
\eea
where $v \in \mathbb{R}^{2,1}$ and $k_0$ is an initial vector (for the hyperboloid, $k_0 = (m,0,0)$ and for the cone $k_0 = (1,1,0)$).

Let $s \mapsto L(s)$ be a UIR of $S_0$, the little group, carried by the Hilbert space $\mathcal{K}$.
The UIR $\chi L$ of $V \rtimes S_0$ carried by $\mathcal{K}$ is then:
\bea
(\chi L)(v,s) = \exp[-i<k_0;v>] L(s).
\eea

Now, we want to induce a representation of $G = V \rtimes S$ from $\chi L$.
From the coset decomposition, $(v,s) = (0,\Lambda _k)(\Lambda _k^{-1}v, s_0)$ ($\Lambda _k$ is the action on the hyperboloid or the cone) we act on the left part (which represents $\mathcal{O}^*$, the hyperboloid or the cone)
\bea
(v,s) (0,\Lambda _p) = (0, \Lambda _{sp})(\Lambda _{sp}^{-1}v,\Lambda _{sp}^{-1}s\Lambda _p ),
\eea
where $p \in \mathcal{O}^*$.
We obtain the following cocycles:
\bea \label{cocycles}
 h : G \times \mathcal{O}^* \to V \rtimes S_0,  \ \ \
h((v,s),p) =  (\Lambda _{sp}^{-1}v, h_0 (s,p));  \nonumber \\
 h_0 : S \times \mathcal{O}^* \to S_0, \ \ \
h_0 (s,p) = \Lambda _{sp}^{-1}s\Lambda _p .
\eea
The UIR is then written this way:
\bea
\left ( ^{\chi L}U(v,s) \phi \right )(k) = \exp [i<k;v>] L(h_0(s^{-1},k))^{-1} \phi(s^{-1}k).
\eea

\subsection{Coherent states from a semidirect product group}\label{get_cs}

In Sections \ref{cs_hyp} and \ref{cs_cone} we shall obtain  coherent states from the Poincar\'e group using the above representation, following the technique described in \citet{AliAG}, \S 10.3.

Coherent states are generally defined as an overcomplete family of vectors, $\eta_x$, in a Hilbert space, $\mathcal H$, indexed by the points $x$ of a measure space $(X, d\mu)$, with the property that
\bea \label{res_id} \int_X  <\phi | \eta_x > < \eta_x |  \psi>\; d\mu (x) =  <\phi | \psi>,\qquad  \phi,\ \psi \in \mathcal{H}.  \eea
The above condition is known as the {\em resolution of the identity}, for it implies (in the sense of weak convergence) that,
\bea \label{res_id2} \int_X | \eta_x > < \eta_x |\; d\mu(x)  =  I_{\mathcal H}.  \eea
In our case, the space $X$ will be one of the coadjoint orbits of the group and the vectors $\eta_x$ will be generated through the action of the representation operators $^{\chi L}U(v,s)$ of the group on a fixed vector $\eta$. The procedure will become clear when we compute these vectors below.

\section{Hyberboloid}\label{cs_hyp}

The hyperboloidal orbit was discussed in Section \ref{sect_2sheethyp_orbit}.
Here we use that setting to define coordinates and compute the invariant measure in order to induce a representation from which we will compute the coherent states.

\subsection{Coordinates and measure}

We set the following space coordinates on the hyperboloid: $q=k_0 \Lambda ^{-1}$, where $k_0 = (m,0,0)$ and $\Lambda$ is the $SO(2,1)$ part of the group element $g$ used in the coadjoint action which generates the orbit.
We can check that $q_0^2 - q_1^2 - q_2^2 =m^2$ is verified.
We also define the momentum coordinates on the cotangent plane:
$p = q (J\cdot v)$, where $J\cdot v$ is given in \eqref{Jv}.

We compute the prime coordinates from the coadjoint action:
\bea \label{prime_coor}
(q', p') = (q, p) M(g)^{-1} = (q \Hat{\Lambda}  + p \Lambda ^{-1}(J\cdot v), p \Lambda ^{-1}),
\eea
where $M(g)$ is the matrix form of the coadjoint action as given in \eqref{matM}.

Note that, from the definition, $p$ depends on the point $q$ to which it is attached.  We thus need to transform the $p$ coordinate in order to compute the invariants.
We postulate $p = \tilde{p}\Lambda _{q\Lambda}$,
where $\Lambda _{q}$ is a pure boost.
The general form for this boost is:
\bea  \label{pure_boost}
\Lambda _q = \frac{1}{m} \bmat q_0 & q_1 & q_2 \\ q_1 & m + \frac{q_1^2}{m+q_0}  & \frac{q_1 q_2}{m+q_0}
\\  q_2  &  \frac{q_1 q_2}{m+q_0}  & m + \frac{q_2^2}{m+q_0}  \emat ,
\eea
we have got it from \citet{Alietal1996_spinCS}.
We then rewrite the following:
\bea \label{p_tilde_prime}
p'= p \Lambda ^{-1}
 =  \tilde{p}\Lambda _{q\Lambda} \Lambda ^{-1}
  =  \tilde{p}\Lambda _{q\Lambda} \Lambda ^{-1} \Lambda ^{-1}_q \Lambda _q
  =  \tilde{p} R \Lambda _q
= \tilde{p}'  \Lambda _q
\eea
where we have defined $R = \Lambda _{q\Lambda} \Lambda ^{-1} \Lambda ^{-1}_q$.  We can check that $R$ is actually a rotation by applying it to the vertex of the hyperboloid which is stable under the action of $R$.

We want to compute the invariant measure on the hyperboloid in our set of coordinates.
From \eqref{prime_coor}, we write $dq_1'$ and $dq_2'$ replacing $q_0'$ using the constraint $q_0^{'2} - q_1^{'2} - q_2^{'2} =m^2$.
We get that the invariant measure is:
\bea \label{meas_hyp}
d\nu = \frac{dq_1 \wedge dq_2}{q_0}.
\eea
In \eqref{p_tilde_prime}, we have defined $\tilde{p}' =  \tilde{p} R$, $R$ being a rotation, then $d\tilde{p}_1 \wedge d\tilde{p}_2$
is easily seen to be invariant.

Finally, the invariant measure on the whole orbit is:
\bea \label{inv_meas}
d\mu = \frac{d\tilde{p}_1 \wedge d\tilde{p}_2 \wedge dq_1 \wedge dq_2}{q_0}.
\eea

\subsection{Induced representation}

We follow the procedure described in Section \ref{ind_rep_meth} in order to obtain the induced representation.

The UIR $\chi L$ of $V \rtimes S_0$ carried by a Hilbert space $\mathfrak{k}$ is:
\bea
(\chi L)(v,s) = \exp[-i<k_0;v>] L(s). \nonumber
\eea
In this case, $S_0$ being only the rotation, we need a one-dimensional representation.  It is written as $e^{i n\theta}$, where $n\in \mathbb{Z}$.  The Hilbert space is thus $\mathfrak{k} = \mathbb{C}$, because we get a complex phase.

Using the process described in Section \ref{ind_rep_meth},
we then get the following cocycles:
\bea \label{cocycles_hyp}
 h : G \times \mathcal{O}^* \to V \rtimes S_0,  \ \ \
h((x,s),p) =  (\Lambda _{sp}^{-1}x, h_0 (s,p));  \nonumber \\
 h_0 : S \times \mathcal{O}^* \to S_0, \ \ \
h_0 (s,p) = \Lambda _{sp}^{-1}s\Lambda _p .
\eea

We need to compute the cocycles for the inverse group element, we get:
\bea \label{inv_cocycle}
h((v,s)^{-1}, p) = (-\Lambda ^{-1} _{s^{-1}p}s^{-1}v, \Lambda ^{-1}_{s^{-1}p}s^{-1} \Lambda _p),
\eea
where $h_0(s^{-1},p) =\Lambda ^{-1}_{s^{-1}p}s^{-1} \Lambda _p $ is a rotation.
This gives the UIR:
\bea
(\chi L)(h((v,s)^{-1}, p)) = \exp[-i <k_0 ; -\Lambda ^{-1}_{s^{-1}p}s^{-1}v>] L(h_0(s^{-1},p)).
\eea

We now have to rewrite the argument $<k_0 ; -\Lambda ^{-1}_{s^{-1}k}s^{-1}v>$.
First, we specify the action of $\Lambda _k$ in both $\mathbb{R}^{2,1}$ and its dual.
\begin{defn}
If we have $v,\ v_0 \in \mathbb{R}^{2,1}$, 3-column vectors, $k,\ k_0 \in \mathbb{R}^{(2,1)*}$, 3-row vectors, then the $3\times 3$ boost matrix $\Lambda$ acts in the following way:
\begin{itemize}
\item $k_0 \Lambda _k  = k$,
 \ \ \ \ \  $k_0 \Lambda ^{-1} _k = \bar{k}$;
\item $ \Lambda _v v_0 = v$,
 \ \ \ \ \  $\Lambda ^{-1} _v v_0 = \bar{v}$;
\end{itemize}
where $\bar{k} = (k_0, -\kv)$.
\end{defn}
We also need the definition of the dual action in the dual pairing. It is: $ <Ad^\# (g) X_1^* , X_2> = <X_1^* , Ad(g^{-1})X_2> $.
In the case of interest here, we rewrite this as: $< k_1 \Lambda _{ks}; v_2> =<k_1 ; \Lambda ^{-1} _{s^{-1}k} v_2> $.
On the LHS, $k$ is in the dual, while on the RHS, $k$ in the original vector space.
We take the transpose of the argument of $\Lambda$ and the inverse of both group elements $\Lambda$ and $s$.
We can then rewrite the argument as follows:
\bea
<k_0 ; -\Lambda ^{-1}_{s^{-1}k}s^{-1}v> &=& -<k_0  \Lambda _{ks} ; s^{-1}v> \nonumber \\
&=& -< ks ; s^{-1}v>\nonumber \\
&=& -k s s^{-1}v = - k v = -<k;v>.
\eea
The UIR is finally written this way:
\bea
\left ( ^{\chi L}U(v,s) \phi \right )(k) = \exp [i<k;v>] L(h_0(s^{-1},k))^{-1} \phi(s^{-1}k).
\eea
Thus, the  UIR we will be using in the following is:
\begin{equation}\label{ourUIR_hyp}
   \left (^{\chi L} U(v,s)\phi \right )(k)
 = \exp[i<k;v>] \exp[-in\theta(k,s)] \phi(s^{-1}k).
   \end{equation}

We can check that the UIR is not square-integrable on the whole group.  We then need to work on a quotient.

\subsection{Quotient and sections}

In order to have a square-integrable representation, we take the quotient to the phase-space.  We follow the left quotient decomposition:
\bea
(\Lambda, v) = \left (\Lambda _{Rq} , (0,p_1,p_2)^t \right ) \left (R, (a,0,0)^t \right ),
\eea
where $R$ is a rotation.
We redefine $\tilde{q} = Rq$.
We work with $(\Lambda _{\tilde{q}} , (0,p_1,p_2)^t)$ which represents the hyperboloid and the space plane orbit.

We need a section to undo the quotient.
The most simple is the Galilean section:
\bea \label{gal_sect}
\sigma _0 : \Gamma \to G \ , \ \ \ \sigma _0(\q,\p) = ((0,\p), \Lambda _q) .
\eea
We also define the principal section:
\bea \label{princ_sect}
\sigma _{\mathcal{P}} : \Gamma \to G \ , \ \ \ \sigma _{\mathcal{P}} (q,p) = (\Lambda _q p, \Lambda _q).
\eea
This is the section we will use in the following.

\subsection{Coherent states for the principal section}

We perform the computations to obtain coherent states on the hyperboloid following the method outlined in Section \ref{get_cs} and using the principal section.

\subsubsection{Definition of the set of vectors}

We choose a set of vectors (vector-valued functions) ${\bf \eta}$ in the Hilbert space
$\mathcal{H} = \mathbb{C} \otimes L^2(\mathcal{V}_m^+, \frac{dq_1\wedge dq_2}{q_0})$, where $\mathcal{V}_m^+$ is the hyperboloid.
Those vectors are transformed by the UIR \eqref{ourUIR_hyp} in the following way:
\bea
({\bf \eta}  _{\sigma _{\mathcal{P}} (q, p)})(k) & =&(U (\sigma _{\mathcal{P}} (q,  p))  {\bf \eta} ) (k) \nonumber \\
& =&(U (\Lambda _q p, \Lambda _q ) {\bf \eta} ) (k) \nonumber \\
&=& e^{ik\cdot \hat{p}} e^{-i n\theta } {\bf \eta} (\Lambda _q ^{-1}k),
\eea
where $\hat{p} = \Lambda _q p$ and $k$ is an arbitrary point on the hyperboloid.

The formal operator is as follows:
\bea
A_{\sigma _{\mathcal{P}}} =  \int _{\Gamma} |{\bf \eta}  _{\sigma _{\mathcal{P}}(q, p)}><{\bf \eta}  _{\sigma _{\mathcal{P}}(q, p)}| \frac{1}{q_0} d{\bf q} d{\bf p},
\eea
where $\Gamma$ is the four-dimensional orbit (the hyperboloid and the space plane).

The change of coordinate $p \to \hat{p}= \Lambda _q p$ gives a complicated Jacobian which is hard to work with.  Instead, we use the change of coordinate $k \to X(k)$ by rewriting the dot product in the exponential:
\bea
k\cdot \hat{p} = k \cdot (\Lambda _q p)
= k \eta \Lambda _q p
= k \Lambda_q^{-1}\eta p
= (k \Lambda_q^{-1}) \cdot p
= X(k) \cdot p,
\eea
where $\eta = {\rm diag}(1,-1,-1)$ is the metric governing the dot product here.  We also use the fact that $\eta \Lambda _q = \Lambda_q^{-1}\eta$ and define $X(k)=k \Lambda_q^{-1}$.

We write $e^{ik\cdot \hat{p}} = e^{i X(k)\cdot p}$ in the integral and compute the Jacobian for the change of coordinate $k\to X(k)$:
\bea
|J| =\frac{1}{mk_0}(q_0 k_0 -q_1 k_1 -q_2 k_2 )= \frac{1}{mk_0} q\cdot k .
\eea

\subsubsection{Integration of the formal operator}

We want to see under which conditions the formal operator $A_{\sigma _{\mathcal{P}}}$ satisfies the resolution of the identity.  We thus compute the integral:
\bea \label{int_hyp_princ}
I _{\phi,\psi} & =& <\phi| A_{\sigma_{\mathcal{P}}} \psi> .
\eea
After a few computations using the integral definition of the $\delta$ function:
\bea \label{delta_fct}
\delta(\x-\y)=\frac{1}{(2\pi)^2} \int _{-\infty}^{\infty} e^{i{\bf v}\cdot(\x-\y)}d{\bf v} ,
\eea
we obtain that:
\bea
I _{\phi,\psi} & =& \int _{\mathcal{V}_m^+}\phi ^* (k)\mathcal{A} _{\sigma_{\mathcal{P}}} (k)\psi(k)\frac{d{\bf k}}{k_0},
\eea
where
\bea \label{int_operatorAP}
\mathcal{A} _{\sigma_{\mathcal{P}}}(k) &=& (2\pi)^2 \int _{\mathcal{V}_m^+} |{\bf \eta} (\Lambda ^{-1} _q k)|^2  \frac{m}{q \cdot k}  \frac{d{\bf q} }{q_0} .
\eea

\subsubsection{Rewriting of the vector argument}\label{rewrite_eta_hyp}

We need to rewrite $|\eta(\Lambda^{-1}_q k)|^2$ as a function of $q$ in order to perform the integral and evaluate $\mathcal{A} _{\sigma_{\mathcal{P}}}(k)$.

We have the following by definition or simple computation:
\begin{itemize}
\item $ \Lambda _k k_0 = k$,  \ \ \  $\Lambda ^{-1} _k k_0 = \bar{k}$;
\item $\Lambda _k \Lambda _q k_0= \tilde {R}\Lambda _q \Lambda _k k_0$, where $\tilde {R}$ is a rotation;
\item $\Lambda ^{-1}_q = \Lambda _{\bar{q}}$;
\item $\overline {\Lambda _k^{-1}q} = \Lambda _k \bar{q}$.
\end{itemize}
\begin{rem}
The second item expresses the fact that $\Lambda _k \Lambda _q$ applied to $k_0$ and $\Lambda _q \Lambda _k$ applied to $k_0$ differ only by a rotation.
Note that this is not true if applied to some other vector.
\end{rem}

We can thus rewrite:
\bea
|\eta(\Lambda^{-1}_q k)|^2 &=& |\eta(\Lambda ^{-1}_q \Lambda _k (m,0,0)^t)|^2  \nonumber \\
 &=& |\eta(\tilde{R}\Lambda _k\Lambda _{\bar{q}}(m,0,0)^t))|^2  \nonumber \\
 &=& |\eta(\Lambda _k \bar{q})|^2  \nonumber \\
 &=& |\eta(q')|^2 ,
\eea
where we define $q'=\Lambda _k \bar{q}$.
We also set that $|\eta|^2$ is invariant under rotation, that is
$|\eta (Rq)|^2=|\eta (q)|^2$.  This means that it is a function of the $0^{th}$ (time) component only.

We compute the $0^{th}$ component of the argument $q'=\Lambda _k \bar{q}$:
\bea
(\Lambda _k \bar{q})_0 = \frac{1}{m} k\cdot q,  
\eea
where $k\cdot q = k_0q_0-k_1q_1-k_2q_2$.

\subsubsection{Evaluation of the integral}

We return to the evaluation of the integral \eqref{int_operatorAP} for $\mathcal{A} _{\sigma_{\mathcal{P}}}(k)$.
We use the fact that $q'_0 = \frac{1}{m}q\cdot k$ and that $\frac{d{\bf q} }{q_0}=\frac{d{\bf q'} }{q'_0}$ (since this is an invariant measure) to write:
\bea \label{oper_A_qprime}
\mathcal{A} _{\sigma_{\mathcal{P}}}(k) &=& (2\pi)^2 \int _{\mathcal{V}_m^+} |{\bf \eta} (q'_0)|^2 \frac{1}{q'_0} \frac{d{\bf q'} }{q'_0}.
\eea
We recall that $\eta$ is square-integrable and that $q'_0 \geq m >0$.
We can then see that $\mathcal{A} _{\sigma_{\mathcal{P}}}(k)$ is actually a constant with respect to $k$ (it only depends on $q'$).  Then,
\bea
I _{\phi,\psi} & =& \mathcal{A} _{\sigma_{\mathcal{P}}} <\phi | \psi >.
\eea

\subsubsection{Resulting coherent states}

The resulting coherent states are the vectors:
\bea
({\bf \eta}  _{\sigma _{\mathcal{P}} (q, p)})(k)
&=& e^{ik\cdot \Lambda _q p} e^{-i n\theta } {\bf \eta} (\Lambda _q ^{-1}k) .
\eea
They have to be normalized by $\sqrt{\mathcal{A} _{\sigma_{\mathcal{P}}}}$ in \eqref{oper_A_qprime} in order to have the resolution of the identity.

\section{Cone}\label{cs_cone}

The conical orbit was introduced in Section \ref{sect_cone_orbit}.
We define coordinates and compute the invariant measure in order to induce a representation from which we will build a family of coherent states.

\subsection{Coordinates and measure}\label{cone_coor_meas}

We use the natural coordinate $q = (q_0,q_1,q_2)$ on the cone embedded in a three-dimensional space.  It satisfies $q_0^2 - q_1^2-q_2^2 = 0$.
The $p$ coordinate is on a plane cotangent to the cone.
Here, we will use $p=(p_0, p_0, p_2)$, this is the coordinate of the plane obtained in the  computation of the orbit.

Once again, the cotangent plane, hence the $p$ coordinate, is attached to the cone at a point $q$.
In the hyperboloidal case, $p$ was changed to $\tilde{p}$ by a pure boost.
We need an equivalent transformation here.
Therefore, we define $p = \tilde {p} \Lambda _{q\Lambda}$, where $\Lambda _q$ is such that $(1,1,0)\Lambda _q = q$.  We then have $q\Lambda _q^{-1} = (1,1,0)$.
It is possible to obtain a matricial representation of $\Lambda _q$:
\bea  \label{lambda_matrix}
\Lambda _q = \frac{1}{q_0+q_1} \begin{pmatrix}  1+q_0^2+q_0-q_1 & q_0q_1 -1 -q_0 +q_1  &  q_2(1+q_0)  \\  q_0q_1 -1 -q_0 +q_1   &  1+q_1^2+q_0-q_1 &  -q_2(1-q_1)  \\   q_2(1+q_0)   &  -q_2(1-q_1) &  q_0+q_1+q_2^2
\end{pmatrix}.
\eea

We now compute the invariant measure,
from the natural coordinates, we write $q' = q\Hat{\Lambda}$ (note that $q'=q\Hat{\Lambda} + p\Lambda ^{-1}(J\cdot v)$, but we study the measure only on the cone, that is without translations $v$).
The invariant measure is $\frac{dq_1 \wedge dq_2}{q_0}$ when there are no translations.

We also have:
\begin{eqnarray*}
p' = p \Lambda ^{-1}
=\tilde{p}\Lambda _{q\Lambda}\Lambda ^{-1}
= \tilde{p}\Lambda _{q\Lambda}\Lambda ^{-1} \Lambda _q ^{-1}\Lambda _q
=  \tilde{p} R  \Lambda _q
= \tilde{p'} \Lambda _q
\end{eqnarray*}
where we have set $R= \Lambda _{q\Lambda}\Lambda ^{-1} \Lambda _q ^{-1}$ and $\tilde{p'} = \tilde{p}R$.
We can check that $R$ is a rotation on the cone.
It satisfies $(1,1,0)R = (1,1,0)$.
We have $\tilde{p'}=\tilde{p} R$, then $d\tilde{p}_1\wedge d\tilde{p}_2$ is invariant.

Finally, the invariant measure on the cone is:
\bea
d\mu = \frac{dq_1 \wedge dq_2 \wedge d\tilde{p}_1\wedge d\tilde{p}_2}{q_0}.
\eea

We present also another representation of the conical coadjoint orbit.  This is based on the projection of the cone on a plane.  This representation will be useful for the coherent states computations.

We represent an element of the cone by $g=(R_{\theta}, \lambda)$,
where $R_{\theta}$ is from the $SO(2,1)$ rotation $\bmat 1 & 0 \\ 0 & R_{\theta} \emat$ and $\lambda$ is the dilation factor (actually $\lambda = \cosh \gamma + \sinh \gamma$ from the $SO(2,1)$ $y$-boost).
The product of two elements is:
\bea
g g' = (R_{\theta}, \lambda)(R_{\theta '}, \lambda ') = (R_{\theta + \theta'}, \lambda \lambda').
\eea
It is obtained from the matricial product of two elements of a semidirect product group and a direct computation for the rotation and the boost.

We will use the angle $\theta$ and the dilation parameter $\lambda \equiv e ^{\gamma}$ as the coordinates on the cone.

We then have another representation of $\Lambda _q$, the action on the cone.  Besides the matrix given in \eqref{lambda_matrix}, we can use $\Lambda _q \equiv \lambda R_{\theta}$.  It will act on the two-vector $(1,0)$ to take it to the projection of $q$ on a plane, that is $(q_1,q_2)$.

\subsection{Induced representation}

We follow the procedure described in Section \ref{ind_rep_meth} in order to obtain the induced representation.

We associate a unitary character $\chi$ to $V = \mathbb{R}^{2,1}$ in the same way as before:
\bea
\chi (x) = \exp (-i<k_0;x>),
\eea
where $k_0=(1,1,0)$.
Let $s \mapsto L(s)$ be a UIR of $S_0$ carried by a Hilbert space $\mathfrak{k}$.  Here $S_0$ is the translation ($n$ in the Iwasawa decomposition).  Then, $L(s)$ is a one-dimensional unitary representation: $e^{it\rho}$, where $t\in \mathbb{R}$ and $\rho$ is the translation parameter.
The UIR $\chi L$ of $V \rtimes S_0$ carried by $\mathfrak{k}$ is:
\bea
(\chi L)(x,s) = \exp[-i<k_0;x>] e^{it\rho}.
\eea
The Hilbert space is $\mathfrak{k}= \mathbb{C}$.

Now, we want to induce a representation of the Poincar\'e group $G = \RR^{2,1} \rtimes SO(2,1)$ from $\chi L$.
From the coset decomposition, $(x,s) = (0,\Lambda _k)(\Lambda _k^{-1}x, s_0)$ (where $\Lambda _k$ is the transformation on the cone and $s_0=n$) we act on $(0,\Lambda _p)$ which represents the cone $\mathcal{O}^*$:
\bea
(x,s) (0,\Lambda _p) = (0, \Lambda _{sp})(\Lambda _{sp}^{-1}x,\Lambda _{sp}^{-1}s\Lambda _p ).
\eea
We obtain the following cocycles:
\bea \label{cocycles_cone}
 h : G \times \mathcal{O}^* \to V \rtimes S_0,  \ \ \
h((x,s),p) =  (\Lambda _{sp}^{-1}x, h_0 (s,p));  \nonumber \\
 h_0 : S \times \mathcal{O}^* \to S_0, \ \ \
h_0 (s,p) = \Lambda _{sp}^{-1}s\Lambda _p .
\eea
They look the same as for the hyperboloid, but it is because of the notation, actually $\Lambda$ and $S_0$ are different matrices.

The UIR is written this way:
\bea
\left ( ^{\chi L}U(x,s) \phi \right )(k) = \exp [i<k;x>] L(h_0(s^{-1},k))^{-1} \phi(s^{-1}k).
\eea
Again, the appearance is the same as in the hyperboloidal case except that the objects are geometrically different.

Finally, the UIR we will be using is written:
\begin{equation}\label{ourUIR_cone}
(^{\chi L} U(v,s)\phi)(k)
 = \exp[i<k;v>]  e^{-it\rho(k,s)}   \phi(s^{-1}k).
\end{equation}
It is similar to the UIR obtained for the hyperboloid in \eqref{ourUIR_hyp}, but here $k$ is a point on the cone, $t \in \RR$ and $\rho$ is a translation parameter.

We can check that this UIR is not square-integrable on the whole group.  We therefore need to work on the quotient.

\subsection{Quotient and section}

We take the quotient to the phase-space to have a square-integrable representation.  We follow the left quotient decomposition:
\bea
(\Lambda, v) = \left (\Lambda _q , \left (\frac{p_0+p_1}{2}, \frac{p_0+p_1}{2},p_2 \right )^t \right )(n, (t, -t,0)^t).
\eea
Note that in the Iwasawa decomposition $\Lambda _q$ is $ka$ the product of a rotation and a boost and $n$ is a translation.

We define a generalized principal section paralleling the principal section for the hyperboloid:
\bea \label{cone_princ}
\sigma _{\mathcal{P}} : \Gamma \to G, \ \ \ \sigma _{\mathcal{P}} (q,p) = (\Lambda ^{\alpha}_q, \Lambda _q^{\beta}p).
\eea
  We have added some freedom with $\alpha$ and $\beta$ exponents.  We will get a constraint on those exponents when computing the coherent states.

\subsection{Coherent states for the generalized principal section}

We now compute the coherent states with the generalized principal section.  We follow the method described in Section \ref{get_cs}.

\subsubsection{Definition of the set of vectors}

We start with a set of square-integrable vectors $\eta$ in the Hilbert space $\mathcal{H} = \mathbb{C} \otimes L^2(\mathcal{V}^+, \frac{dq_1\wedge dq_2}{q_0})$, where $\mathcal{V}^+$ is the cone.  We transform them using the UIR given in \eqref{ourUIR_cone}:
\bea
(\eta _{\sigma _{\mathcal{P}}})(k) = e ^{i <k;\hat{p}>}e^{-it\rho}\eta(\Lambda _q^{-\alpha}k),
\eea
where $\hat{p} =\Lambda _q^{\beta} p $ and $k$ is an arbitrary point on the cone.

The formal operator is defined as follows:
\bea A _{\sigma _{\mathcal{P}}}  = \int _{\Gamma} |\eta _{\sigma _{\mathcal{P}}} > <\eta _{\sigma _{\mathcal{P}}} | \frac{d\q d\p}{q_0}. \eea

\subsubsection{Integration of the formal operator}

In order to compute the integral of the formal operator $I_{\Phi,\Psi} = <\Phi | A _{\sigma _{\mathcal{P}}} \Psi >$, we use again the integral definition of the $\delta$ function given in \eqref{delta_fct}.
We also work out the Jacobian of the change of coordinate $p \to \hat{p} = \Lambda _q^{\beta} p$.  To do this, we use the projection coordinates defined in Section \ref{cone_coor_meas}.
We rewrite: $\hat{p} =  \Lambda _q^{\beta} p \equiv \lambda ^{\beta} R_{\beta \theta} \p$, where $\p$ is the projection of $p$ on the punctured plane.
Then, we have that the Jacobian is $|J| = \lambda ^{2\beta}$.

When rewriting the integral with this Jacobian, we obtain:
\bea
I_{\Phi,\Psi} &=& \int _{\mathcal{V}^+} \Phi ^* (k) \mathcal{A} _{\sigma_{\mathcal{P}}} (k) \Psi (k) \frac{dk}{k_0},
\eea
where
\bea \label{a_sigma_cone}
\mathcal{A} _{\sigma_{\mathcal{P}}} (k) &=& (2\pi)^2 \int _{\mathcal{V}^+}  |\eta (\Lambda _q^{-\alpha}k)|^2 \frac{1}{\lambda ^{2\beta}} \frac{1}{k_0}\frac{d\q}{q_0} .
\eea

\subsubsection{Rewriting of the vector argument}

In order to rewrite the argument of $\eta$,
we use the projection setting and the isomorphism of the cone with the punctured plane.

We redefine the point $k$ to be the initial point $(1,0)^t$ on which the rotation $R_{\phi}$ and the dilation $\tau$ act, that is $k\to {\bf k} = \tau R_{\phi} (1,0)^t$.
We also have that $k_0 = \tau$, the time component only depends on the dilation.  (Note that we can also obtain this from $k_0^2 = k_1^2 + k_2^2 = \tau ^2 \cos ^2 \phi + \tau ^2 \sin ^2 \phi$.)
Similarly, we write $\Lambda _q$ as $\lambda R_{\theta}$, then $\Lambda _q^{-\alpha} = \lambda^{-\alpha} R_{-\alpha\theta}$.

With all this information, we are able to write:
\bea
|\eta (\Lambda _q^{-\alpha}k)|^2 &=& |\eta (\lambda^{-\alpha} R_{-\alpha\theta} \tau R_{\phi} (1,0)^t)|^2 \nonumber \\
&=& |\eta (\lambda^{-\alpha} \tau R_{\phi-\alpha\theta} (1,0)^t)|^2 \nonumber \\
&=& |\eta (\lambda^{-\alpha} \tau  (1,0)^t)|^2 ,
\eea
where we have considered that $|\eta|^2$ is rotational-invariant.

We rewrite the integral \eqref{a_sigma_cone}:
\bea
\mathcal{A} _{\sigma_{\mathcal{P}}} (k) &=& (2\pi)^2 \int _{\mathcal{V}^+}  |\eta (\lambda^{-\alpha} \tau  (1,0)^t)|^2 \frac{1}{\lambda ^{2\beta}} \frac{1}{\tau}\frac{d\q}{q_0} .
\eea
We can see that $|\eta|^2$ depends only on the length of the vector $\lambda^{-\alpha} \tau$, we thus change the variable to $q'_0=q'_1=\lambda^{-\alpha} \tau$ (and $q'_2=0$).

\subsubsection{Evaluation of the integral}

We recall that the measure $\frac{d\q}{q_0} $ is invariant.
Also, provided $2\beta = -\alpha$, we have $\tau \lambda ^{2\beta} = q'_0$ too.
We get the resolution of the identity since the operator is now written:
\bea \label{A_sigma_cone}
\mathcal{A} _{\sigma_{\mathcal{P}}} &=& (2\pi)^2 \int _{\mathcal{V}^+}  |\eta (q'_0)|^2 \frac{1}{q'_0}\frac{d\q'}{q'_0}
\eea
and does not depend on $k$.  Also, recalling that $\eta$ is square-integrable and $q'_0 >0$, for an initial vector $\eta$ in the domain of the unbounded operator, multiplication by $\frac{1}{q'_0}$, we have:
\bea
I_{\Phi,\Psi} &=& \mathcal{A} _{\sigma_{\mathcal{P}}} \int _{\mathcal{V}^+} \Phi ^* (k) \Psi (k) \frac{dk}{k_0} \nonumber \\
&=& \mathcal{A} _{\sigma_{\mathcal{P}}} <\Phi | \Psi >.
\eea

\subsubsection{Resulting family of coherent states}

The vectors $\eta _{\sigma _{\mathcal{P}}}$ are coherent states for the section:
\bea
\sigma _{\mathcal{P}} (q,p) = (\Lambda ^{-2\mu}_q, \Lambda _q^{\mu}p),
\eea
that is:
\bea
(\eta _{\sigma _{\mathcal{P}}})(k) = e ^{i <k;\Lambda _q^{\mu}p>}e^{-it\rho}\eta(\Lambda _q^{2\mu}k)
\eea
are a family of coherent states for $\mu \in \mathbb{Z}$ and a suitable $\eta$
with the normalization by $\sqrt{\mathcal{A} _{\sigma_{\mathcal{P}}}}$ given in \eqref{A_sigma_cone}.  This is a promising new result.

The parameter $\mu$ appearing in the computation may have a physical interpretation.  This is something to be explored.

\section*{Conclusion}

 Our aim in this paper was to study the nature of all the coadjoint orbits of the Poincar\'e group in $2+1$ dimensions. We did this here by working with a matricial form of the group. From there, we have computed and classified the coadjoint orbits as well as the orbits of $SO(2,1)$ acting on the dual of $\RR^{2,1}$.  We have obtained a degenerate orbit, the two-sheeted hyperboloid, the upper and lower cones and the one-sheeted hyperboloid; they all also appear as two-dimensional coadjoint orbits.  Moreover, the hyperboloids and the cones appear with their cotangent planes as four-dimensional coadjoint orbits.  We have also seen that the representation generating orbits and the four-dimensional coadjoint orbits were linked. While some of the results presented here also appear elsewhere, the complete classification that we have done, is new. 

 Subsequently, we  used the information obtained to define coordinates and an invariant measure on two of the  orbits, the hyperboloid and the cone.  We have also induced a representation and computed the coherent states on each of them.
For the hyperboloid, we have obtained coherent states for the principal section.  For the cone, we have obtained a family of coherent states for a generalized principal section.

  There are two areas of study to which we would like to apply these results in future publications: $(1)$ computing wavelet-like transforms on the cone and the hyperboloid and their coadjoint orbits, which would be useful in signal analysis; $(2)$ the physical problem of an electron moving in a constant magnetic field, normal to a cone and a hyperboloid and the resulting Landau levels.

\end{document}